\documentclass[10pt]{article}

\usepackage{amsmath}
\usepackage{amssymb}

\usepackage{graphicx}

\usepackage{cite}

\usepackage{color}

\usepackage[labelfont=bf,labelsep=period,justification=raggedright]{caption}

\makeatletter
\renewcommand{\@biblabel}[1]{\quad#1.}
\makeatother

\date{}

\usepackage{paralist}

\newcommand{\mean}[1]{\left\langle #1 \right\rangle}

\begin{document}

\begin{flushleft}
{\Large
\textbf{Redistribution spurs growth by using a portfolio effect on risky human capital}
}
\\
Jan Lorenz$^{1,2,\ast}$,
Fabian Paetzel$^{3}$,
Frank Schweitzer$^{1}$
\\
\bf{1} Chair of Systems Design, ETH  Z\"urich, Kreuzplatz 5, 8032 Z\"urich, Switzerland
\\
\bf{2} Center for Social Science Methodology, Carl von Ossietzky
  Universit\"at Oldenburg, Ammerl\"ander Heerstr. 114--118, 26129
  Oldenburg, Germany \footnote{Present address: Jacobs University Bremen, Campus Ring 1, 28759 Bremen, Germany}
\\
\bf{3} Center for Social Policy Research, Universit\"at Bremen, Mary-Somerville-Str. 5, 28359 Bremen, Germany
\\
\end{flushleft}

\section*{Abstract}
We demonstrate by mathematical analysis and systematic computer simulations that redistribution can lead to sustainable growth in a society. In accordance with economic models of risky human capital, we assume that dynamics of human capital is modeled as a multiplicative stochastic process which, in the long run, leads to the destruction of individual human capital. When agents are linked by fully-redistributive taxation the situation might turn to individual growth in the long run. We consider that a government collects a proportion of income and reduces it by a fraction as costs for administration (efficiency losses). The remaining public good is equally redistributed to all agents. Sustainable growth is induced by redistribution despite the losses from the random growth process and despite administrative costs. Growth results from a portfolio effect. The findings are verified for three different tax schemes: proportional tax, taking proportional more from the rich, and proportionally more from the poor. We discuss which of these tax schemes performs better with respect to maximize growth under a fixed rate of administrative costs, and with respect to maximize the governmental income. This leads us to some general conclusions about governmental decisions, the relation to public good games with free-riding, and the function of taxation in a risk taking society.

\section*{Introduction}

This paper shows how redistribution of income spurs growth of human capital in a society just because of a portfolio-effect. Our model captures the concept of ``risky human capital'' from recent economic literature (see \cite{Grochulski.Piskorski2010Riskyhumancapital}), where human capital is described by a multiplicative stochastic process which, in the long run, leads to the destruction of individual human capital. We model the random growth or decline of human capital as proportional to the individual endowment of income. We couple agents by fully-redistributive taxation \cite{Meltzer.Richard1981RationalTheoryof} (that means collected taxes are equally redistributed) which is associated with efficiency losses \cite{Persson.Tabellini1994IsInequalityHarmful}. Redistribution re-balances gains and losses from individuals and works as a portfolio effect which spurs growth into the individually lossy stochastic processes.

Economic literature does not explicitly point out the portfolio effect through redistribution in the relationship between inequality and growth of human capital. So far, in the politico-economic literature, three basic reasons are discussed of why redistribution is beneficial for society. The first branch of literature stresses the insurance aspect through redistribution (see \cite{Mirrlees1971ExplorationinTheory}). Mirrlees highlights that from a welfare maximizing point of view, the level of redistribution should be at a level on which the poor do not suffer and both the poor and the rich have an incentive to improve their situation \cite{Mirrlees1971ExplorationinTheory}.\footnote{Mirrlees motivates the insurance aspect with fairness-considerations. Therefore, this literature is also partly linked to the literature about social preferences \cite{Fehr.Schmidt1999TheoryOfFairness, Bolton.Ockenfels2000ERCTheoryof}. In a series of experiments, it is shown that subjects have a preference for avoiding high degrees of inequality even if they have to resign payoff.}. More recently, the literature of socio-political unrest \cite{Barro2000InequalityandGrowth,Forbes2000ReassessmentofRelationship} points out that redistribution guarantees social stability and reduces the effort the society has to make when inequality is high.

In the second branch, redistribution reduces the disincentive for the poor for taking too high risks \cite{Aghion.Bolton1997TheoryofTrickle-Down}. Redistribution increases the endowment of the poor. The poor reduce the demand for loans and invest more efficient which means that they take less risk. The efficiency in the economy is improved, growth will be higher. Redistribution spurs growth because of decreased disincentives in society.

The third branch describes the transmission channel of the median-voter-approach between inequality and growth \cite{Persson.Tabellini1994IsInequalityHarmful}. The utility maximizing calculus of the median-voter determines the level of redistribution. The median voter proposes the median level of redistribution. This level is unbeatable in pairwise majority decisions. If the inequality is high, the median-voter enforces a high level of redistribution. Redistribution is assumed to induce efficiency losses, as in our model. High levels of inequality are associated with high efficiency losses and therefore low growth rates. For instance in \cite{Perotti1996Growthincomedistribution} it is shown that high degrees of inequality are not associated with high levels of social spending. This motivated the modification and extension of this approach by the consideration of institutions \cite{Acemoglu.Johnson.ea2005InstitutionsasFundamental} and elites \cite{Glaeser.Scheinkman.ea2003injusticeofinequality}.

In contrast to above mentioned work on the relationship between inequality and growth, we argue that neither socio-political nor incentive considerations have to be taken into account to show that redistribution spurs growth. Our model excludes incentive-incompatibilities, voting-approaches and normative concepts like insurance or fairness considerations, to point out the effectiveness of the pure portfolio effect through redistribution.

As we will see, the effect of combining lossy proportional stochastic growth and linear lossy redistribution is non-trivial as both processes lead to destruction of human capital when they run independently, while their combination can enable survival. This ``magic'' effect of induction of growth from two lossy processes is based on the portfolio effect known from investment science \cite{Luenberger1998InvestmentScience}: Gains and losses are re-balanced by redistributing income into human capital ``assets'', which ensures optimal growth of the portfolio. The effect has been discussed before under different names, such as repeated Kelly games, Kelly optimal portfolio, and re-balancing of asset allocations, and seems to be rediscovered from time to time in a new context \cite{Kelly1956newinterpretationof,
  Bouchaud.Mezard2000Wealthcondensationin,
  Malcai.Biham.ea2002Theoreticalanalysisand,
  Marsili.Maslov.ea1998Dynamicaloptimizationtheory,
  Medo2009Breakdownofmean-field,
  Medo.Pismak.ea2008Diversificationandlimited,
  Slanina1999possibilityofoptimal,
  Slanina2004Inelasticallyscatteringparticles,
  Yaari.Solomon2010Cooperationevolutionin}.
It's applicability in various fields is laid out in \cite{Yaari.Stauffer.ea2009IntermittencyandLocalization}. We will call the phenomenon \emph{portfolio re-balancing effect} in the
following. This interprets each subject with its human capital endowment as an asset and the society as the portfolio. Asset values change stochastically, and taxation and redistribution of income re-balance the values of the different assets.

The econophysics literature has studied extensively the statistical mechanics of money based on its conservation \cite{Dragulescu.Yakovenko2000Statisticalmechanicsof,Chakraborti.Chakrabarti2000Statisticalmechanicsof,Patriarca.Chakraborti.ea2006Influenceofsaving} and within a kinetic exchange model \cite{Chatterjee.Chakrabarti2007Kineticexchangemodels} (see also the paper and the literature review in  \cite{Yakovenko.Rosser2009ColloquiumStatisticalmechanics}). The topic of taxation and redistribution has been introduced in these models \cite{Guala2009Taxesinsimple}. Regarding this stream of literature, we focus on the growth enhancing effect of redistribution neglecting conservation of money.

The purpose of this paper is to discuss the role of the portfolio re-balancing effect of fully-redistributive taxation for the growth of society's human capital. By systematic computer simulation we quantify conditions of the stochastic growth process, and the taxation schemes that prevent the destruction of the accumulated human capital of the society. Further on, we quantify optimal tax rates which maximize growth of society's human capital, and how a selfish government optimizes its income.

\section*{Methods}

We describe our method by the following steps:
\begin{compactitem}
 \item Specification of the outlined economic model on the endogenous variables human capital $h$ and income $y$ in terms of a straight forward transition of human capital to income, redistribution of income between agents and independent random multiplicative production of human capital proportional to income. Exogenous variables are: tax rate $a$, rate of administrative cost $b$, taxation scheme, the distribution of random growth factors, and the number of agents $N$.
 \item Theoretical demonstration what conditions are most interesting for studying the effect of portfolio re-balancing on growth.
 \item Theoretical demonstration of dynamics at border cases.
 \item Presentation of some example runs of the process.
 \item Description of the setup of a systematic simulation the extraction of the average growth factor from simulation data.
\end{compactitem}

\subsection*{Specification of the economic model}

Let us consider a society of $N$ agents, each of which is characterized at time $t$ by
its \emph{human capital} $h_{i}(t)$ which is a positive scalar value. Consequently, $h(t) \in \mathbb{R}^N$ is the \emph{human capital vector} of all $h_{i}(t)$. Let us denote the \emph{total human capital} at time $t$ by $H(t)=\sum_{i=1}^N h_i(t)$.

The production process is: human capital is used to produce \emph{income} $y_i$ income is taxed and fully-redistributed, and income is directly invested in human capital. This is formalized by the equations:
\begin{align}
 y_i(t) &=  \mathrm{prod}_i(h_i(t))   \label{eq:system1}, \\
 y_i(t+1) &= \mathrm{redis}_i(y(t))   \label{eq:system2}, \\
 h_i(t+1) &= \mathrm{HCprod}_i(y_i(t+1)) \label{eq:system3}.
\end{align}
Notice that $\mathrm{prod}_i$ and $\mathrm{HCprod}_i$ are functions operating on individual values, while $\mathrm{redis}_i$ is scalar-valued but takes the whole income vector as input.
These scalar-valued functions serve as component functions for the vector-valued functions $\mathrm{prod}$, $\mathrm{HCprod}$, and $\mathrm{redis}$ which are self-maps on $\mathbb{R}^N$.

Given an initial human capital vector $h(0)$ the evolution of human capital in the system of individuals is described by the equation
\begin{equation}
 h(t+1) = \mathrm{HCprod}(\mathrm{redis}(\mathrm{prod}(h(t)))). \label{eq:system_h}
\end{equation}

Consequently, the evolution of income is given by
\begin{equation}
 y(t+1) =  \mathrm{redis}(\mathrm{prod}(\mathrm{HCprod}(y(t)))). \label{eq:system_y}
\end{equation}
We call $Y(t) = \sum_{i=1}^N y_i(t)$ the \emph{total income}. Growth (positive or negative) of this aggregated variable is analyzed in the following. The total income evolves of course tightly related to the evolution of the total human capital $H(t)$ because of Eq.~(\ref{eq:system3}).

\paragraph{Production of income and human capital}
We assume that production is directly transfered into income
\begin{equation}
 y_i = \mathrm{prod}_i(h_i) = h_i,  \label{eq:production}
\end{equation}
which means that the wage is equal to one. The production of human capital is assumed to be based on an individual multiplicative stochastic event
\begin{equation}
  h_i =  \mathrm{HCprod}_i(y_i) = \eta_i(t) y_i
  \label{eq:HCproduction}
\end{equation}
where $\eta_i(t)$ is a realization of the positive random variable $\eta$. If agent $i$ at time $t$ has income $y_i(t)$ then after producing its human capital is $\eta_i(t)y_i(t)$. When $\eta_i(t)<1$ human capital declines, otherwise it grows. Thus, we assume that human capital for the next round of production is build from current income times a random factor. This can be interpreted as a generation model where each generation lives for one period and invests its income in the human capital of its successor. In an innovative economy the value of old human capital quickly dissolves and new human capital must constantly be produced by investing income. Thus, the human capital production function is also reasonable on shorter time scales then generations.

Without subscript $i$, $\mathrm{prod}$ and $\eta(t)$ are meant as vectors. Thus, growth dynamics of human capital vectors $h$ reads $\mathrm{HCprod}(y) = \eta(t) y$. The product $\eta(t) y$ is meant component-wise, $\eta(t)$ being an equally sized vector of independent realizations of $\eta$.

Our production function has only the input factor human capital and is therefore simplified. The simplification is motivated by the idea of the human capital intensive production in modern economies which is in line with endogenous growth theory (see \cite{LucasJr.1988mechanicsofeconomic}). For reasons of comparability, we present in the following some feasible extensions of our model. A standard Cobb-Douglas production function also includes the input of \emph{capital} $k$ and \emph{labour} $l$ (accompanied by their exponents $0 < \alpha,\beta < 1$ with $\alpha+\beta=1$). Moreover, the transfer of income into human capital is accompanied by \emph{consumption} $c$ and \emph{saving} rates $s$. This would imply Equation (\ref{eq:production}) to look as
\begin{equation}
 y_i = \mathrm{prod}(h_i) = h_i k^\alpha l^\beta  \nonumber
\end{equation}
and Equation (\ref{eq:HCproduction}) to be
\begin{equation}
  h_i = \mathrm{HCprod}(y_i) = \eta_i(t)(1-s)(1-c)y_i. \nonumber
\end{equation}
Equation (\ref{eq:system_y}) would read
\begin{equation}
 y(t+1) =  \mathrm{redis}(\;\eta_i(t)(1-s)(1-c)y_i k^\alpha l^\beta \;).
\end{equation}
When $s,c,k,l,\alpha,\beta$ are all constants the term $\eta_i(t)(1-s)(1-c) k^\alpha l^\beta$ is a draw from a random variable from the distribution as $\eta$ but scaled by a constant factor $(1-s)(1-c) k^\alpha l^\beta$. Modification of these factors have thus the same effect as a multiplicative scale in the random variable $\eta$. By holding $k$ constant, we implicitly assume that savings are equal to depreciation of capital. Our model is thus based on a standard economic growth model and is simplified to focus on the \emph{portfolio re-balancing effect}.

\paragraph{Redistribution}
We quantify the redistribution function with three different taxation schemes: proportional taxation, a progressive scheme where agents have to pay everything above a dynamically chosen maximal tax-free income, and a regressive scheme where agents have to pay either a dynamically chosen fee -- like a per capita premium -- or all their income if they cannot afford the full fee. (No worries, agents get back some income because of the redistribution.) All three schemes we specify by the same two independent parameters: the \emph{tax rate} $a$, which determines the fraction withdrawn from the total income of all agents, and the \emph{rate of administrative cost} $b$, which determines the fraction withdrawn by the government from the raised taxes before redistribution to agents.

Let us call the amount of taxes collected from agent $i$ to be $\mathrm{tax}_i(y)$. Notice, that it depends on the vector of income. This enables us to define dynamically adjusted taxation schemes which take the distribution of income into account. Naturally, $\mathrm{tax}_i(y)$ should be confined between zero (no tax) and $y_i$ (tax equals income).

The tax revenue is collected by a government at a central place, which involves administrative costs (efficiency losses cf. \cite{Persson.Tabellini1994IsInequalityHarmful}),
which are assumed to be proportional to the amount of taxes raised, i.e.
$b\in[0, 1]$ denotes the \emph{rate of administrative cost}. Consequently, the \emph{public good} for redistribution is the raised taxes minus the cost:
\begin{equation}
  \mathrm{pg}(y) = (1-b)\sum_{i=1}^N\mathrm{tax}_i(y), \label{eq:pg}
\end{equation}
while the \emph{government income} is
\begin{equation}
  \mathrm{gi}(y) = b\sum_{i=1}^N\mathrm{tax}_i(y). \label{eq:govincome}
\end{equation}
Because of a fully-redistributive tax, $\mathrm{pg}(y)$ is divided with equal shares among all agents, i.e.~for every agent income increases by an amount $\mathrm{pg}(y)/N$. Other mechanisms of redistribution in a related model are analyzed in \cite{Guala2009Taxesinsimple}.
The redistribution function (net income) for agent $i$ is thus
\begin{equation}
  \mathrm{redis}_i(y) = y_i - \mathrm{tax}_i(y) +
  \frac{\mathrm{pg}(y)}{N}.
  \label{eq:redis}
\end{equation}
Depending on the position within society, an agent could be a net tax payer or a net transfer recipient. We specify the tax function $\mathrm{tax}_i(y)$ with respect to the tax rate $a\in[0, 1]$ such that it holds
\begin{equation}
 \sum_i    \mathrm{tax}_i(y) = a \sum_i y_i = aY. \label{eq:atax}
\end{equation}

We distinguish three schemes of taxation which differ in from whom the fraction $a$ of the total income is raised: (i) proportionally from everyone, (ii) more than proportionally from the poor (regressive), or (iii) more than proportionally from the rich (progressive). In progressive and regressive taxation schemes tax rates differ in given income brackets. We consider extremal cases to pronounce differences.

\begin{compactenum}[(i)]
\item \emph{Proportional tax} is the classical taxation scheme where each
  agent has to pay a fraction $a$ of its individual income
  \begin{equation}
    \mathrm{tax}_i(y) = ay_i.
    \label{tax:prop}
  \end{equation}
\item \emph{Regressive tax} charges a fixed fee $c_\mathrm{fee}(y)>0$ from
  everyone if possible, otherwise all income is charged.
  \begin{equation}
    \mathrm{tax}_i(y) = \min\{x_i, c_\mathrm{fee}(y)\}
    \label{eq:dynfee}
  \end{equation}
  In the latter case, the agent still receives its proportion from the
  public good, so it will not be without income after redistribution. The fee has to be such that (\ref{eq:atax}) is fulfilled for the current income vector. (It is easy to see that this is possible and unique.)
\item \emph{Progressive tax} charges all income exceeding a threshold of tax free income \mbox{$c_{\max}(y)>0$} from everyone. Every agent with income below
  $c_{\max}(y)>0$ pays no taxes.
  \begin{equation}\mathrm{tax}_i(y) = \max\{y_i-c_{\max}(y),
    0\}\label{eq:dynmax}
  \end{equation}
  The threshold has to be determined such that (\ref{eq:atax}) is met. (It is easy to see that this is possible and unique.) Notice,
  that the income of an agent who has to pay taxes is larger than $c_{\max}(y)$
  after redistribution because of its share from the public good.
\end{compactenum}

The three schemes are comparable in that the total amount of raised taxes is always a fraction $a$ of the total income, regardless of the shape of the distribution of income. Thus, they all deliver a public good of  $\mathrm{pg}(y) = (1-b)a Y$. Note, that in the related model of \cite{Guala2009Taxesinsimple} taxes are raised when agents trade and not every time period from every one.

The regressive and the progressive tax assign different tax rates in two different tax brackets $[0,c(y)]$ and $[c(y),\infty]$. The regressive tax scheme (where $c=c_\mathrm{fee}$) taxes 100\% in the lower tax bracket and 0\% in the upper. The progressive tax scheme (where $c=c_{\max}$) taxes 100\% in the upper tax bracket and 0\% in the lower. Figure \ref{fig:1} demonstrates an example with six agents of different income. It is shown what will be charged from each agent and how income looks after redistribution for each of the three schemes.

Note, that the dynamic fee $c_{\mathrm{fee}}(y)$ and the dynamic maximum $c_{\max}(y)$ are implicitly defined, to meet the condition of Eq.~(\ref{eq:atax}). In realistic taxation systems, it might seem impractical to determine the fee and the maximum after the current income of all agents is known. In reality, one would only adjust thresholds for the next turn. We omitted that modification to prevent delay effects. Probably, this modification would cause only minor changes.

\subsection*{On the distribution of random human capital growth factors}

Analysis of human capital production without redistribution ($a=0$ in Eq.~(\ref{eq:system2})) does not involve interaction. Therefore, it is enough to focus on a single agent and Eq.~(\ref{eq:system_y}) collapses to $y(t+1) = \eta(t)y(t)$. Let $\eta$ have finite variance. With $y(0)=1$ it holds
\begin{equation}
  y(t+1) = \eta(t)y(t) = \eta(t)\eta(t-1)\cdots\eta(1)\eta(0) = \prod_{s=0}^t\eta(s). \label{eq:prod}
\end{equation}
This resembles the human capital life cycle model \cite{Grochulski.Piskorski2010Riskyhumancapital} where human capital $h_t$ at time $t$ is determined as the result of a stochastic process $h_t = \sigma_{t-1}\cdots\sigma_1\theta i$ where $i$ is the initial investment in human capital, $\theta$ is a stochastic shock to the human capital investment and $\sigma_s$ are stochastic human capital depreciation shocks.

Eq.~(\ref{eq:prod}) is equivalent to
\begin{equation}
  \label{eq:1b}
  \log y(t+1) = \log\eta(t) + \log y(t) = \sum_{s=0}^t\log\eta(s).
\end{equation}
The central limit theorem applied to (\ref{eq:1b}) implies that
the distribution of the random variable $\log y(t)$ gets closer and
closer to a normal distribution $\mathcal{N}(\mu_t, \sigma_t)$ with mean and variance parameters
\begin{equation}
  \mu_t = t\mu_{\log\eta}\;;\quad
  \sigma_t^2 =  t\sigma_{\log\eta}^2
  \label{eq:mean-t}
\end{equation}
with $\mu_{\log\eta} = \mean{\log\eta}$ and  $\sigma_{\log\eta}^2 = \mean{(\log\eta)^2} -  \mean{\log\eta}^2$.
Consequently, for $t\to\infty$ the distribution of $y(t)$ approaches the log-normal distribution $\log\text{-}\mathcal{N}(t\mu_{\log\eta}, \sqrt{t}\sigma_{\log\eta})$. Based on that fact, we chose the log-normal distribution with its two characterizing parameters as the distribution of $\eta$ in our simulation setup.

The expected value of income might grow, while every individual trajectory of $y(t)$ dies out. The condition for this seemingly contradictory situation is
\begin{equation}
  \mu_{\log\eta}<0<\log\mu_\eta \label{eq:effectcondition}
\end{equation}
which is equivalent to $\mean{\eta}_\mathrm{geo}=\exp\mu_{\log\eta}<1<\mu_\eta = \mean{\eta}$ i.e.~the arithmetic mean of $\eta$ being larger than one, while its geometric mean is less than one. Elementary explanations of this effect are given in
\cite{Luenberger1998InvestmentScience,
  Yaari.Solomon2010Cooperationevolutionin,
  Redner1990RandomMultiplicativeProcesses}.
It can be shown that for long enough time span any single trajectory grows only with the geometric mean $\mean{\eta}_\mathrm{geo}$.
For log-normal distributions of $\eta$ the two inequalities in (\ref{eq:effectcondition}) are equivalent to $\mu_{\log\eta}$ being negative while $\sigma^2_{\log\eta}$ is sufficiently large $\sigma^2_{\log\eta} > -2\mu_{\log\eta}$. 

This situation forms the basis for the effect of growth which is induced by coupling lossy multiplicative stochastic growth with lossy redistribution in a finite population. Redistribution helps the system to realize a growth rate somewhere in between the geometric and the arithmetic mean of $\eta$.

To reduce the number of independent parameters in simulation, we choose log-normal distributions where $\mean{\eta}\cdot\mean{\eta}_\mathrm{geo} = 1$ holds. Under this condition, the two parameters of the log-normal distribution $\mu$ and $\sigma$ (mean and standard deviation of the underlying normal distribution) are represented by one free parameter which allows for different skewness, but keeps the balance of the expected value $\mean{\eta}$ an the realized growth rate $\mean{\eta}_\mathrm{geo}$. This condition enables that destruction and growth are theoretically possible for distributions of this class.

\subsection*{Theoretical analysis of border cases and an example}

We are interested in the average realized growth rate and its dependence on the independent parameters. For some border cases we can theoretically derive that average growth is exponential $Y(t+1) = gY(t)$ and also quantify the magnitude of the growth rate $g$.

\textbf{Case 1: } Only redistribution, no stochastic production of human capital ($\mean{\eta}_\mathrm{geo}=1=\mean{\eta}$). For any taxation scheme, and any $N$ it holds $g=(1-ab)$. Thus, there is never growth of human capital.

\textbf{Case 2:} Only stochastic production of human capital but no redistribution ($a=0$). For any taxation scheme, any $b$ and any $N$ it holds $g=\mean{\eta}_\mathrm{geo}$.

\textbf{Case 3:} 100\% tax and infinite number of agents ($a=1$, $N=\infty$). All trajectories act as one, and for any taxation scheme $g=(1-ab)\mean{\eta}$. The growth with the mean can be realized because in an infinite society no rare but large growth event is ``missing''.

Based on the last two cases we argue, that for intermediate $a$ and finite $N$ the average growth rate lies somewhere in between, but we do not have an analytic expression for it.

Figure~\ref{fig:2} gives an example, where we fix  $\mean{\eta}=3/2$ and $\mean{\eta}_\mathrm{geo}=2/3$. This implies log-normal parameters $\mu=-0.405$ and $\sigma=1.274$, which we use as the distribution of intermediate risk in simulation. Under this distribution income declines with a probability of 62.5\%, it at least doubles with probability 19.4\%, and it will be more than ten times
larger with probability 1.7\%. Tax and admin rates are set at intermediate levels $a=0.3$, $b=0.2$. Trajectories are computed according to \eqref{eq:system_y} for a society of $N=10$ agents, each starting with human capital equal to one. Trajectories are shown for all three tax schemes. Each trajectory is computed with the same random draws from the random variables $\eta_i(t)$ for each $i$ and $t$, to allow for a direct comparison of the different taxation schemes.

In Figure~\ref{fig:2} progressive taxation (iii) leads to the largest growth of total income, proportional taxation (i) resulted also in a growing society while regressive taxation (ii) fluctuates between growth and decline with no clear trend visible. When there was no redistribution at all there would be decline (with a growth factor $\mean{\eta}_\mathrm{geo}=0.667$). Pure redistribution without production of human capital would also imply decline (with a growth factor of $1-ab=0.94$). The performance ranking progressive better than proportional better than regressive holds even if we vary the essential parameters admin rate $b$, tax rate $a$ and the size of the population $N$ as shown in the lower part of Figure~\ref{fig:2}. Regarding their impact on the dynamical behavior, we see that for a higher admin rate ($b=0.6$) growth might turn to decline. A lower tax rate ($a=0.01$), can also imply destruction of income and human capital for all three taxation schemes. In this case the portfolio re-balancing effect of redistribution is not used well and this is not compensated by the savings from the lower loss of redistribution ($1-ab=0.998$). Finally, in a larger society with $N=100$ all schemes achieve larger growth factors.

Progressive taxation may contribute to disincentives for the decision to invest income into the production of human capital. Especially in our extremal case income above a maximal income is taxed by 100\%. To that end, let us assume that every agent can decide what fraction of its income to invest while the remaining income's value remains as is. Taxation and redistribution remains obligatory. If the agent is to maximize the value of its income after human capital production, production and redistribution, the rational decision would be to invest all, as long as the expected value of human capital is larger than the invested capital. This holds also for progressive taxation, as any increase of expected value before redistribution increases the expected value of the public good and thus the own expected value after redistribution. In our model, progressive taxation does not remove the rationality of the choice of investing all into production of human capital, but makes difference to not investing in human capital smaller. We do not touch the issue of free-riding (which would be to avoid paying taxes) in the above argument. When free-riding was possible, not paying taxes but receiving a share from the public good, is of course rational under every taxation scheme.

\subsection*{Simulation setup, independent and dependent variables}

For each of the three taxation schemes we aim to get an overview about the dependence of the average growth factor on the tax rate and the rate of administrative costs. Further on, we want to control for the effect of society's size and lognormal distributions which are more or less risky (in the sense of higher or lower right-skewness).

Consequently, we set up a systematic computer simulation to estimate the \emph{average growth factors} $g$. Table \ref{tab:simsetup} shows the list of independent variables, their ranges in the simulation setup. Growth factors are estimated on the basis of stochastic trajectories of total income $Y(t)$ for 500 time steps by regressing $\log(g)$ in $\log Y(t) = \log N + t\cdot\log(g)$. The intercept in the regression is naturally fixed at $\log N$. From 100 estimated growth rates of such stochastic trajectories we compute the average growth factor as the geometric mean. The geometric mean is used because it better fits the central tendency of the distribution, as a growth rate is naturally a parameter larger than zero an thus typically log-normally distributed. Changing to the arithmetic mean would shift the results a bit, as the arithmetic mean is always larger than the geometric mean.

To get on overview about the parameter space we cover the $(b,a)$-plane by a fine grid while the number of different sizes of the society and different distributions was kept low to make computations finish within less than two weeks on a laptop. See the \texttt{matlab}-code in the supporting material which produces the simulation data (see function \texttt{dataMSPgrowthrates}).

As our focus is on maximizing growth factors, optimal tax rates, and government income let us define further variables which depend on the average growth factor as a function of $a$ and $b$, $g(b,a)$: For a fixed admin rate $b$ we define the \emph{maximal growth factor} $g_{\max}(b) = \max_a g(b,a)$ and the \emph{optimal tax rate} $a_\mathrm{opt}(b) = \mathrm{argmax}_ag(b, a)$. (Note, that $\mathrm{argmax}$ is not necessarily unique, but empirical results support the conjecture that there is only one local maximum, and consequently $a_\mathrm{opt}(b)$ is unique.) The \emph{rate of government income} at the current total income is $\mathrm{gov}(b,a) = b\,a\,Y(t+1)/Y(t) = b\,a\,g(b,a)$. The \emph{rate of government income under optimal tax rate} as a function of $b$ is defined by $\mathrm{gov}_\mathrm{opttax}(b) = b\,a_\mathrm{opt}(b)\,g_{\max}(b)$.

\section*{Results}

\subsection*{Description of figures}

\begin{figure}[ht!]
\begin{center}
    \includegraphics{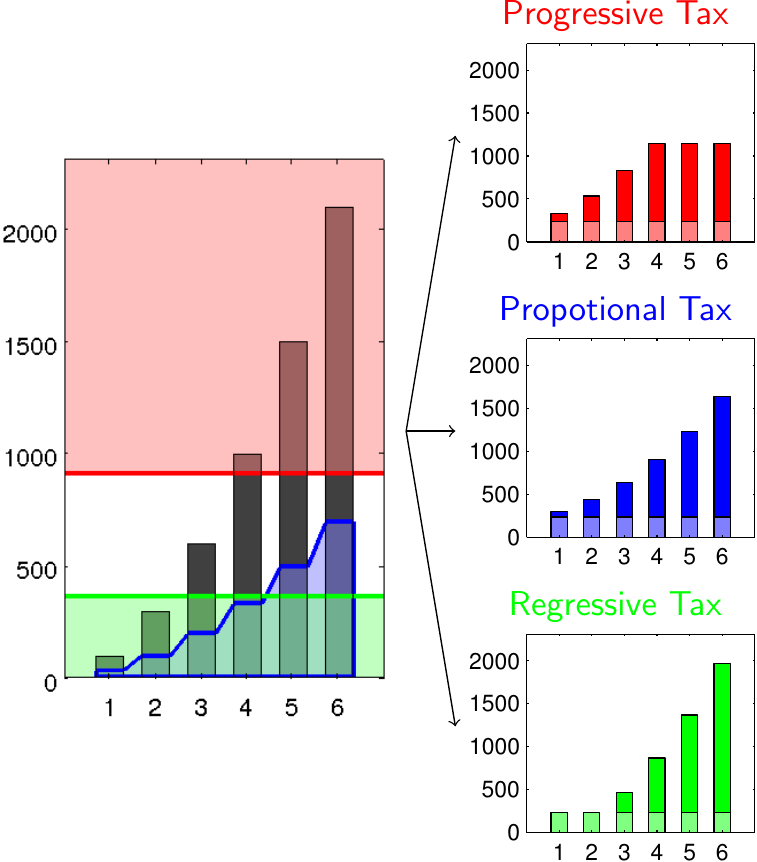}
\end{center}
\caption{{\bf Demonstration of the redistribution function.} \emph{\color{blue}Proportional tax}, \emph{\color{green}regressive tax} and \emph{\color{red}progressive tax} functions. Six agents with
    income $y=[100, 300, 600, 1000, 1500, 2100]$. All tax functions are
    such that $a=1/3$ of the total income ($=5600$) is taxed, the
    administrative cost is set to $25\%$ ($b=0.25$). In numbers: $\mathrm{pg}(y) = 1400$, which implies $c_\mathrm{fee}=366\frac{2}{3}$ and
    $c_\mathrm{max}=911\frac{1}{9}$.} \label{fig:1}
\end{figure}

\begin{figure}
\begin{center}
    \includegraphics{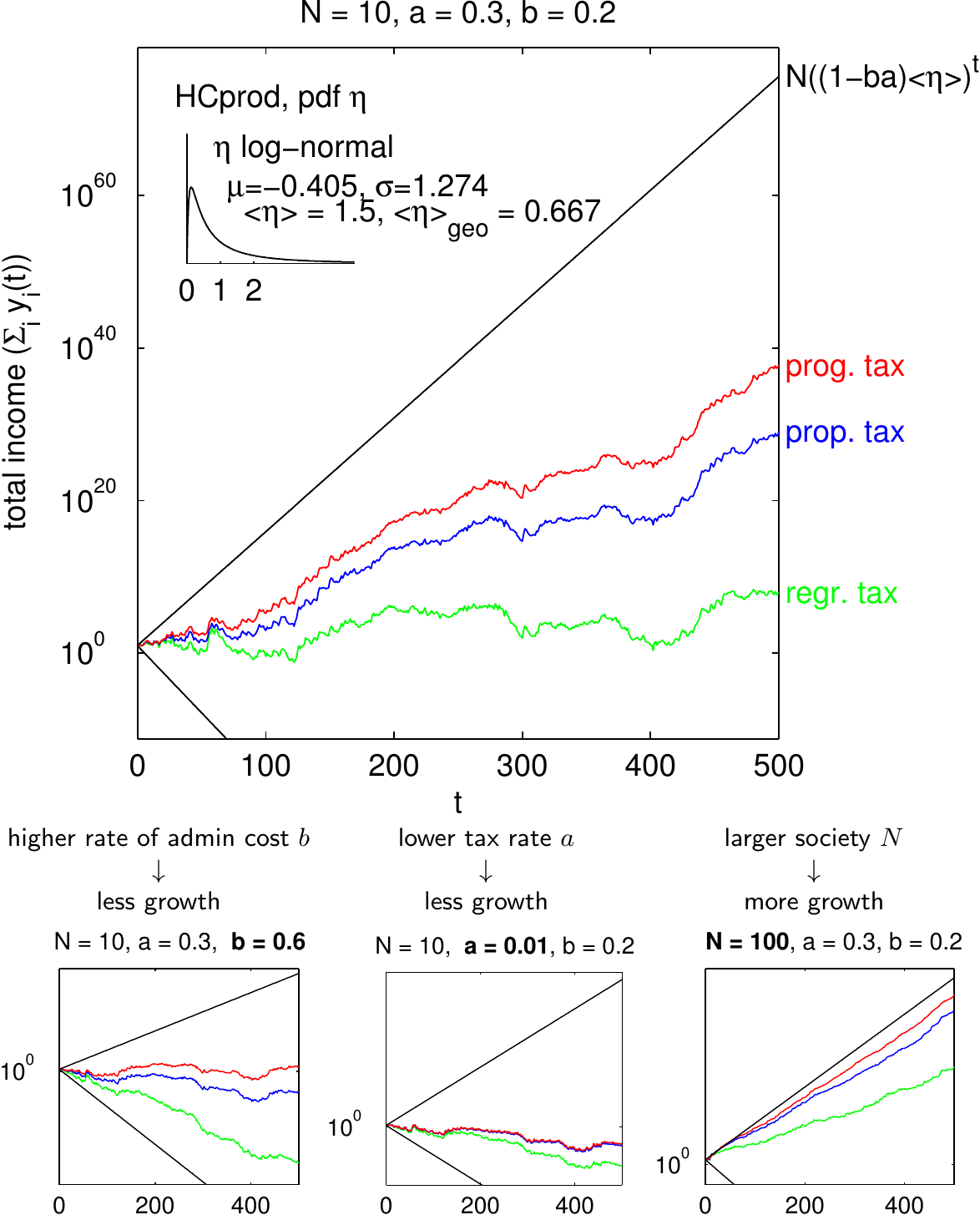}
\end{center}
  \caption{\textbf{Example trajectories for tax schemes and exemplary modifications in subfigures. } Trajectories computed with the same realizations of the random variables $\eta_i(t)$ and iteration of Eq. (\ref{eq:system_y}). Black lines show limiting cases: lower line shows ``no tax'' $Y(t) = N{(\mean{\eta}_\mathrm{geo})}^t = 10\cdot(0.667)^t$; upper line shows ``full tax and infinite number of agents'' $Y(t) = N{((1-ab)\mean{\eta})} = 10\cdot(1.41)^t$. }\label{fig:2}
\end{figure}

\begin{figure}
  \centering
    \includegraphics[width=12cm]{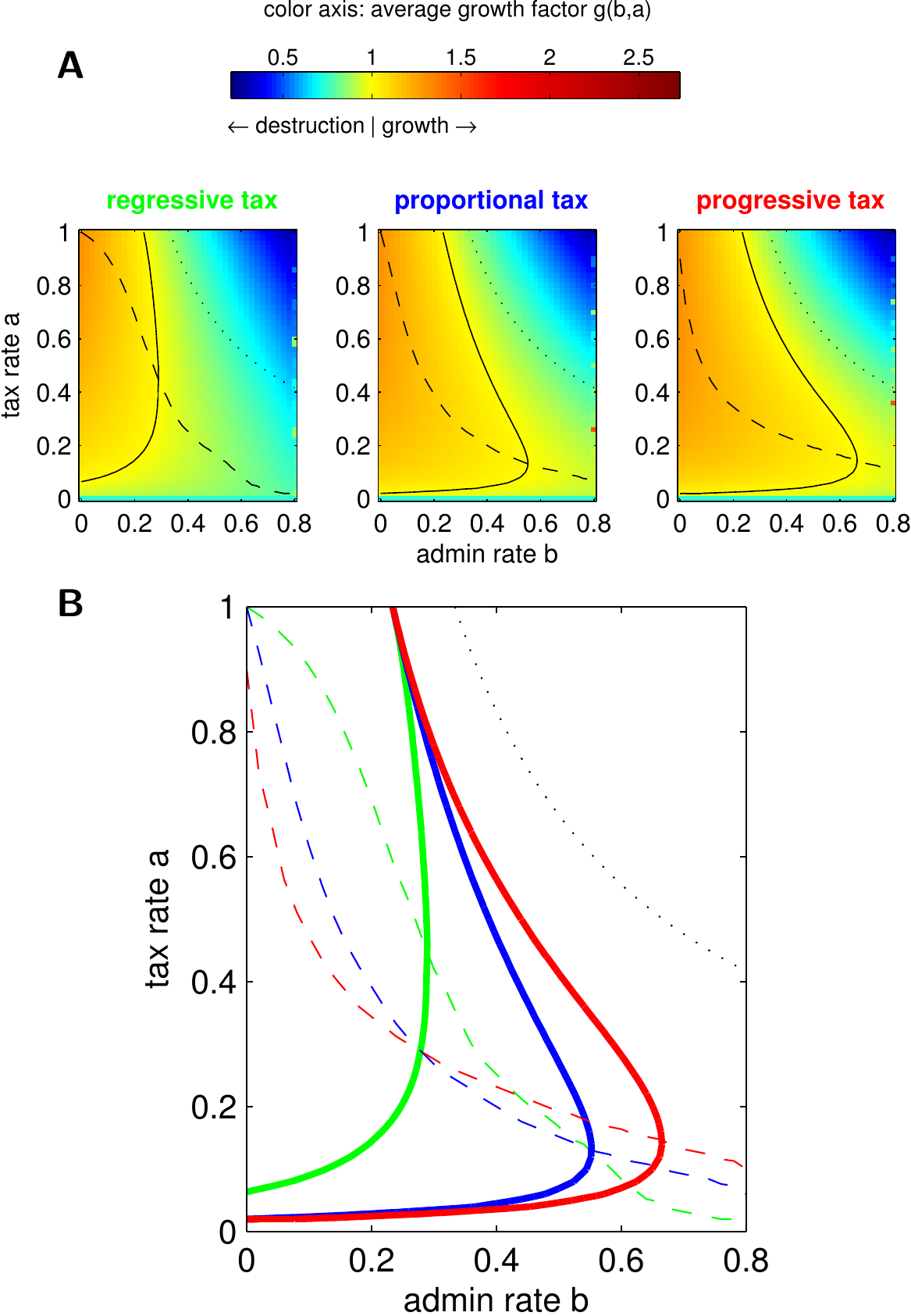}
  \caption{\textbf{Zones of growth and destruction.} Based on the average growth factor $g(b,a)$ for $N=10$, $\mean{\eta}=1.5$,
    and $\mean{\eta}_\mathrm{geo}=0.667$ (cf. Figure~\ref{fig:2}). \textbf{A} $g(b, a)$ in the $(b, a)$-plane color-coded as specified in color bar for all three taxation schemes. Solid lines divide zones of income growth from income destruction. Dashed lines are optimal tax rates for given admin rate $a_\mathrm{opt}(b)$. Above the dotted line income destruction must happen. \textbf{B} Lines of A in one plot for comparison. Colors indicate tax
    schemes. Linestyles as above.}\label{fig:3}
\end{figure}

\begin{figure}
  \centering
    \includegraphics[width=0.38\columnwidth]{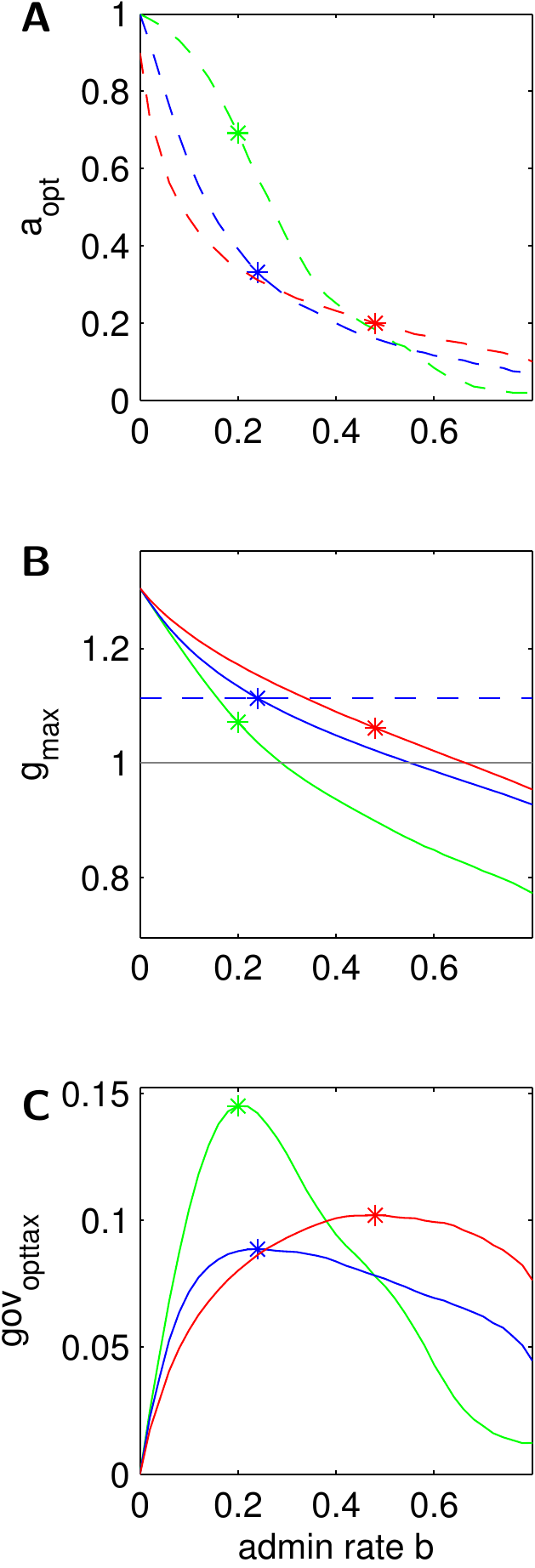}
  \caption{\textbf{Optimal tax rate, maximal growth rate and maximal income of government.} \textbf{A} Optimal tax rate $a_{\mathrm{opt}}(b)$, \textbf{B} maximum
    achievable growth rate $g_{\mathrm{max}}(b)$, \textbf{C} governmental
    income rate under optimal tax rate $\mathrm{gov}_\mathrm{opttax}(b)$ for different tax
    schemes: proportional tax (blue), regressive tax (green), progressive tax (red). Stars indicate the location of the maximum of $\mathrm{gov}_\mathrm{opttax}$ in all plots. The dashed line is to show that under the maximal $\mathrm{gov}_\mathrm{opttax}$ proportional taxation gives the largest growth factor. Parameters as in \ref{fig:3}.}\label{fig:4}
\end{figure}

Figures \ref{fig:1}--\ref{fig:4} are the core Figures to understand the message of our paper. Figures \ref{fig:1} and \ref{fig:2} explain and demonstrate our economic model and Figures \ref{fig:3} and \ref{fig:4} show main results.

We visualize our simulation results exemplarily for $N=10$ and the distribution of intermediate risk $(\mean{\eta},\mean{\eta}_\mathrm{geo})=(1.5,0.667)$ in Figure \ref{fig:3}. We show in Figure 3\textbf{A} the average growth factor
$g(b,a)$ color-coded in the $(b,a)$ parameter plane for each of the three
tax schemes. In each plot, the solid line divides the ``zone of sustainable growth of income'' (yellow to red) from the ``zone of income destruction'' (yellow to blue). The dashed line shows the optimal growth maximizing tax rate $a_\mathrm{opt}(b)$ for a given admin rate. In Figure \ref{fig:3}\textbf{B}, we show the critical lines dividing the zones of growth and destruction and the optimal tax rates in one plot to compare the three taxation schemes. (In all plots the black dotted line shows the maximally possible size of the zone of growth, where $(1-ba)\mean{\eta} = 1$. Above this line it is trivial that income can not grow.

Figure \ref{fig:4}\textbf{A} shows the maximal growth factor $g_{\max}(b)$, \ref{fig:4}\textbf{A} the optimal tax rate $a_\mathrm{opt}(b)$, and  \ref{fig:4}\textbf{C} the rate of government income at optimal tax rate $\mathrm{gov}_\mathrm{opttax}(b)$. All are functions of the admin rate $b$ and they are shown for all three taxation schemes in our standard color-code.

\begin{figure}
   \includegraphics[width=\textwidth]{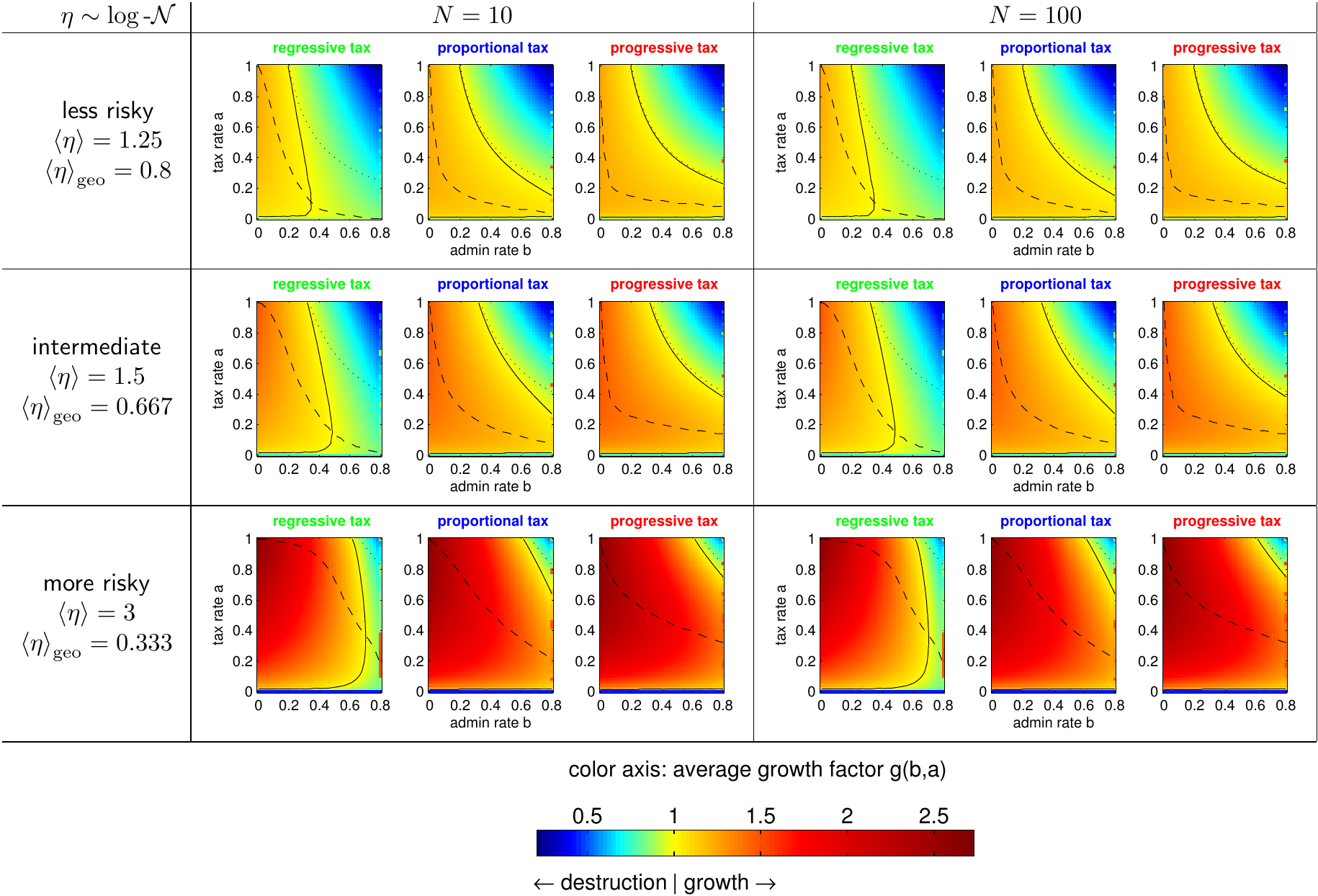}
  \caption{\textbf{More agents, less and more risky human capital.} Extension of Figure
    \ref{fig:3}\textbf{A} with less and more risky human capital production functions (in rows) and more agents (in another column).}\label{fig:5}
\end{figure}

\begin{figure}
\begin{center}
   \includegraphics[width=0.85\textwidth]{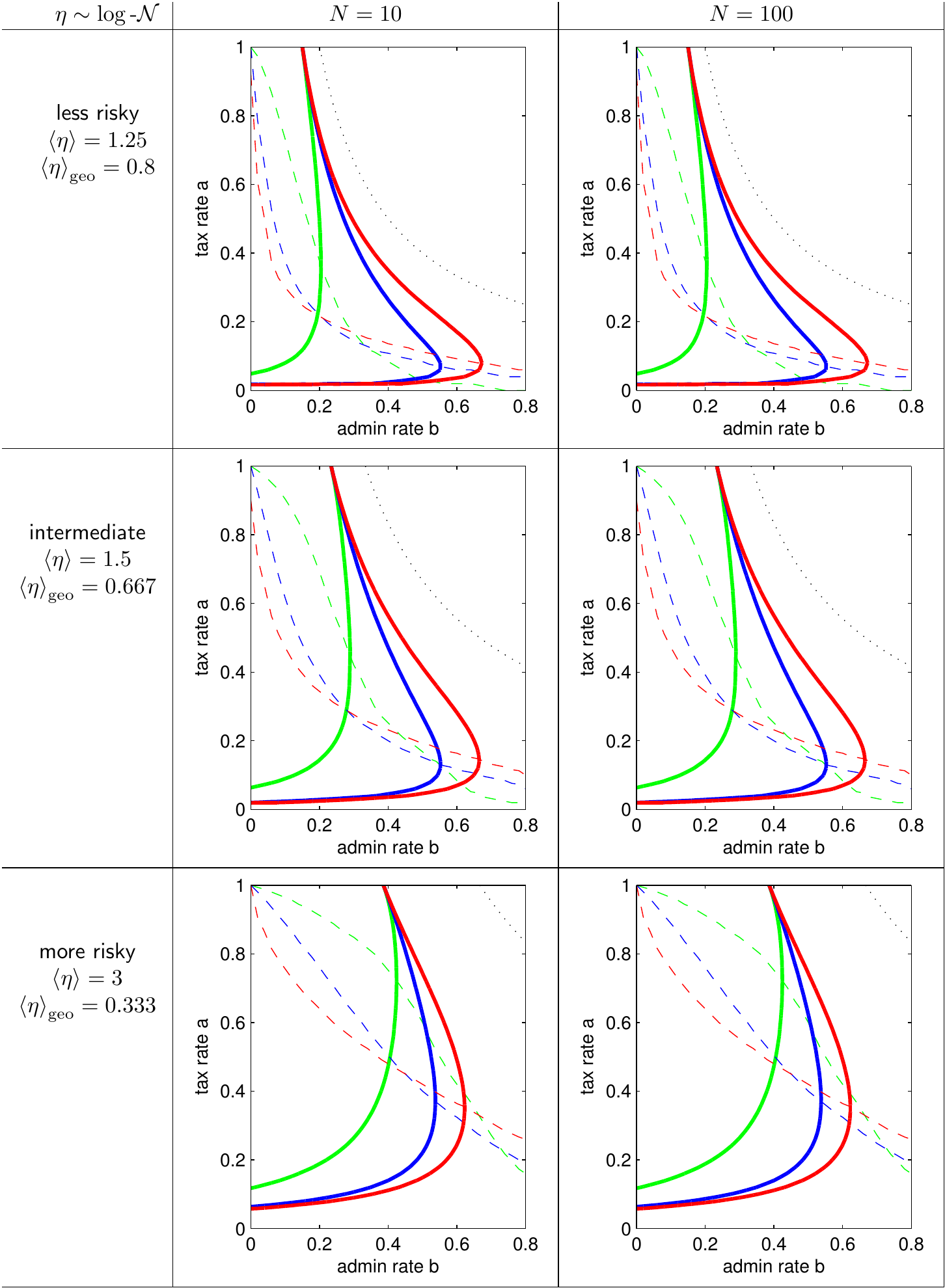}
\end{center}
\caption{\textbf{Zones of growth and destruction for more agents, less and more risky human capital.} Simulation results analog to Figure
    \ref{fig:3}\textbf{B} with less and more risky human capital productions functions (in rows) and more agents (in another column). (Some lines as in Figure
    \ref{fig:7}.)}\label{fig:6}
\end{figure}

\begin{figure}
   \includegraphics[width=\textwidth]{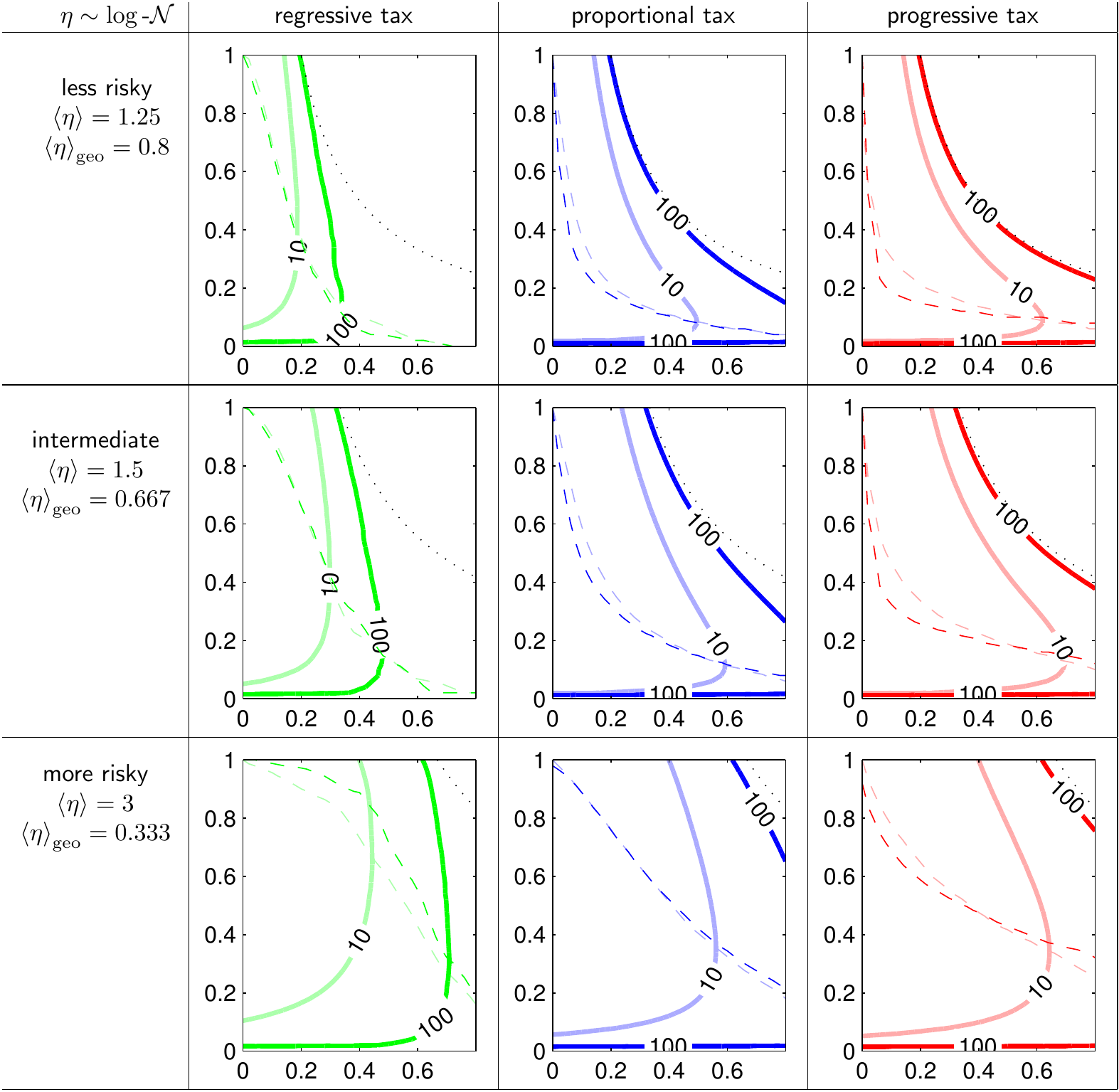}
  \caption{\textbf{Zones of growth for different population sizes.} Same lines as in Figure \ref{fig:6}, but such that different population sizes are in one plot and taxation schemes in columns. }\label{fig:7}
\end{figure}

\begin{figure}
   \includegraphics[width=\textwidth]{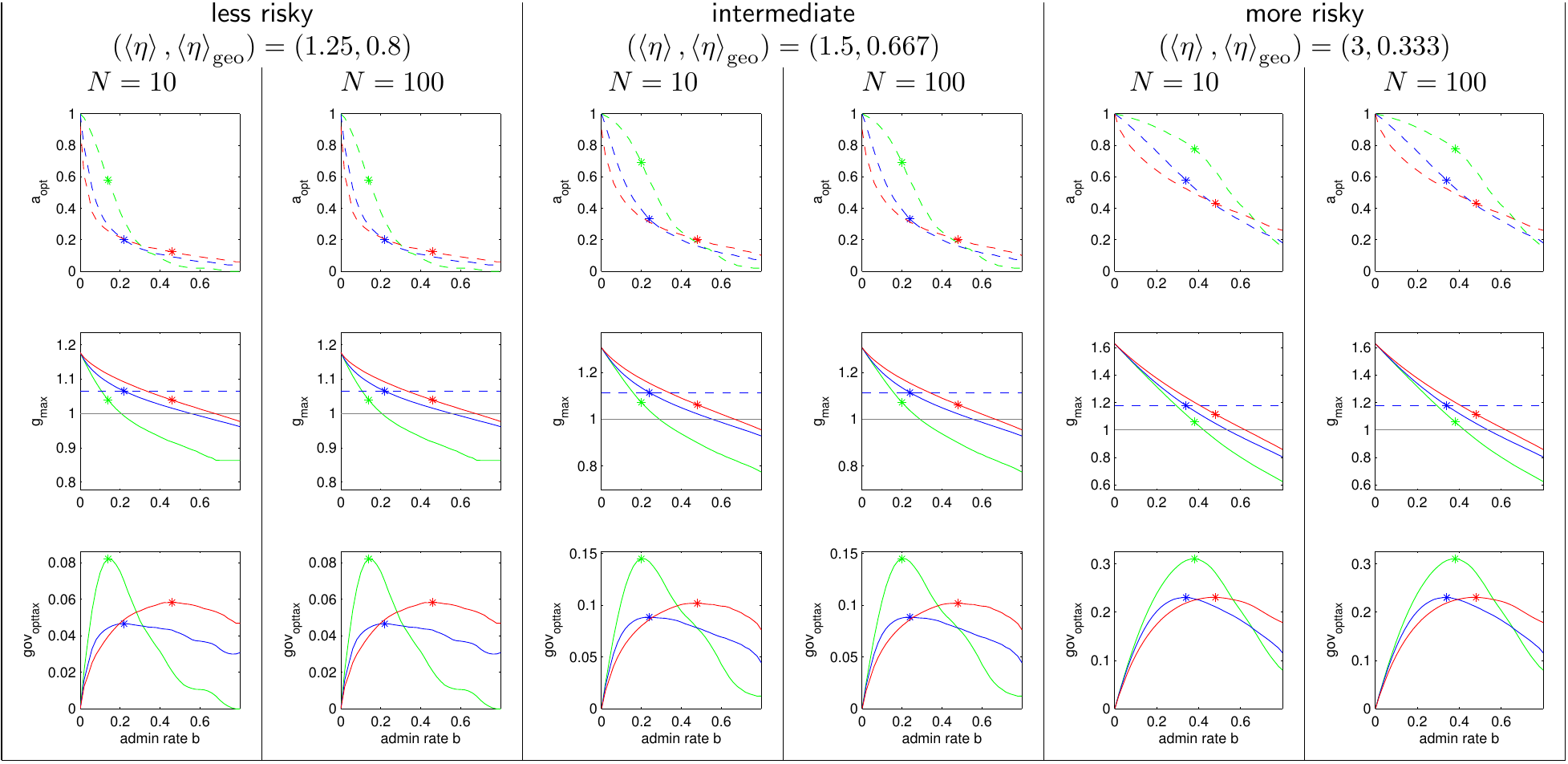}
  \caption{\textbf{More agents, less and more risky human capital and its impact on optimal tax rate, maximal growth rate and maximal income of government.} Simulation results analog to Figure
    \ref{fig:4} with less and more risky human capital production functions and more agents. }\label{fig:8}
\end{figure}

Figure \ref{fig:5}--\ref{fig:8} are extensions of Figures \ref{fig:3} and \ref{fig:4} by showing also the data for $N=100$, a less risky, and a more risky distribution of $\eta$. Figures \ref{fig:5}, \ref{fig:6}, and \ref{fig:8}, are the pure extensions of Figures \ref{fig:3}\textbf{A}, \ref{fig:3}\textbf{B}, and \ref{fig:4}, while Figure \ref{fig:7} is a regrouping of lines to different subplots focusing on comparison of $N=10$  and $N=100$.

\subsection*{On growth and destruction of income and human capital}

Figure \ref{fig:3} and related Figures \ref{fig:5}, \ref{fig:6}, and \ref{fig:7} summarize what combination of tax rates and admin rate allow for society's income and human capital to grow sustainable. It is interesting to focus on situations where either the tax rate $a$ or the admin rate $b$ is fixed:

\textbf{Constant tax rate $a$:} Raising the admin rate $b$ always lowers the growth factor $g(a,b)$ and turns the growth regime at some point into the destruction regime.

\textbf{Constant admin rate $b$:} The average growth factor $g(a,b)$ is not monotonic in $a$. For very high and very low tax rates, the growth factor is the lowest and can lead to income destruction, while only intermediate tax rates prevent this. High tax rates tend to lower the growth factor because a larger fraction of the total income is reduced by the admin rate (see the definition of the public good (\ref{eq:pg})). On the other hand, very low tax rates lower the growth factor because the portfolio re-balancing effect is not used optimal, thus part of the human capital ``gambles it self away''.

This characterization is ubiquitous for all taxation schemes, both population sizes and the riskier and less risky distribution of $\eta$. It holds ubiquitous that the zone of sustainable growth of regressive taxation is contained in the zone of growth of proportional taxation, which is further contained in the zone of growth of progressive taxation, when $N$ and the distribution of $\eta$ are kept constant. When the taxation scheme and the distribution of $\eta$ is kept constant the zone of growth of the smaller society ($N=10$) is always contained in the zone of growth of the larger society ($N=100$).

These findings suggest that the portfolio re-balancing effect is used more effectively, when the society is large and when proportionally more is taken from the rich (progressive tax), than from the poor (regressive tax). Simple inclusions of zones of growth do not hold for comparisons of low risk, intermediate and risky distribution of $\eta$.  We refrained from comparing them in detail, because our balancing condition $\mean{\eta}\cdot\mean{\eta}_\mathrm{geo}=1$ is somehow arbitrary.

In the following we answer five questions about optimal choices of tax and admin rates for different perspectives.

\subsection*{On growth maximizing tax rates and taxation schemes}

\textbf{(a)} \emph{What is the optimal tax rate $a_{\mathrm{opt}}(b)$ and how does it differ between the three taxation schemes?} The optimal tax rate is 100\% under admin rate $b=0$ for any tax system but it declines fast with rising admin rate as can be seen in Figure~\ref{fig:4}\textbf{A}. Within the range $0<b<0.35$ the progressive tax scheme reaches the lowest optimal tax rate, regressive taxation the highest. For a realistic admin rate of about 20\% the optimal tax rate in the progressive tax scheme and the proportional tax scheme is less than 30\%, but it is larger than 50\% under the regressive tax scheme. The ranking is inverted for large admin rates. From Figure \ref{fig:8} it can be seen that these rankings and the switch of the ranking also holds for larger societies being more drastic for low admin rates and less drastic for high admin rates. For riskier societies the optimal tax rates are larger in general.

\textbf{(b)} \emph{Which taxation scheme reaches the largest average growth factor for a given admin rate and optimal choice of the tax rate?}  As can be seen in the central panel of Fig.~\ref{fig:4}\textbf{B} the progressive tax scheme achieves the highest maximal growth factors for all admin rates. The proportional tax scheme is always second and the regressive tax scheme ranks last. Hence, the largest growth factor is reached with the progressive tax scheme that takes more than proportionally from the rich.

\subsection*{On income maximizing governments}

Let us assume, that governments are forced to choose tax rates close to the optimal tax rate. A government might be forced to do so in an informed and democratic society, when the impact on the tax rate on average growth is known and when voters wish that growth rates of total income or human capital are as large as possible. Under this assumption the rate of government income under optimal tax rate $\mathrm{gov}_\mathrm{opttax}$ is of interest, because we can ask what admin rate a government might choose to maximize its income. The rate of administrative costs is usually also under the control of the government, but we assume that a government is not forced to optimize it for growth. Optimal would of course be to have no administrative cost. One reason is that it might be easier for a government to argue, that the administrative cost cannot be lowered, because of fixed contracts. Another argument is that an alternative party which could overtake government has possibly the same interest of increasing its income and consequently democratic competition does not work as easy. Based on these two assumption and simulation results we can answer three further questions:

\textbf{(c)} \emph{Which admin rate $b$ would a self interested government choose?}
Figure~\ref{fig:4}\textbf{C} shows that the rate of government income under optimal tax rates is not monotonic in $b$. In particular, for higher values of $b$ the growth of total income becomes smaller, hence even a government maximizing its income has no incentives to raise the admin rate to the largest possible. This is because large admin rates reduce growth. The admin rate where the government income is maximal is marked by ``$\ast$'' in all three panels. It varies with the taxation scheme: lowest admin rates for regressive tax, highest admin rates for progressive tax. This ranking only changes for the riskier distribution of $\eta$ (see Figure \ref{fig:8}), with proportional tax having lowest admin rates.

\textbf{(d)} \emph{Which taxation scheme would a self interested government choose?}
Looking at absolute values of $\mathrm{gov}_\mathrm{opttax}(b)$ a self interested government would choose the regressive tax because it gives the maximum income of all schemes, even at moderate admin rates. Thus, the largest government income is reached with a scheme that takes more than proportional from the poor. This is caused mainly due to the fact that the optimal tax rate is much larger under the regressive tax scheme. Consequently, the share raised by the admin rate is also larger as under other schemes. Thus, this result crucially depends on the assumption that a government is forced to implement the optimal tax rate, while the taxation scheme is given.

\textbf{(e)} \emph{Which taxation scheme delivers the largest average growth factors under a self interested government?}
Looking at the ``$\ast$''-symbols in Figure \ref{fig:4}\textbf{B} which come from optimal admin rates with respect to  $\mathrm{gov}_\mathrm{opttax}(b)$ in Figure \ref{fig:4}\textbf{C}, we find that proportional tax delivers the highest average growth factors. This holds although regressive taxation can deliver the highest income for the governemt as seen in \textbf{(d)}. The reason is that the regressive tax has much lower growth rates than the other schemes in general. It also holds although progressive taxation always delivers higher growth rates than the other regimes as seen in \textbf{(b)}. The reason here is that the progressive tax attracts the government to raise the admin rate to optimize its income.

\section*{Discussion}

In our view, redistribution enhances the dynamic potential for human capital production of an economy. The enhancement can be explained by the effectiveness of the portfolio effect. 
The enhancement seems is possible for proportional, regressive and progressive taxation schemes.

The answers to questions \textbf{(b)}, \textbf{(d)}, and \textbf{(e)} deliver three different choices of one of the three taxation schemes. \textbf{(b)} suggest that the progressive taxation scheme should be chosen because it is always superior when a rate of administrative cost is fixed. Consequently, taking proportionally more from the rich is socially optimal when we can assume that the rate of administrative costs is an externally fixed parameter. \textbf{(d)} instead suggests, that a selfish government would decide for the progressive taxation scheme because under this scheme growth optimizing tax rates are much higher which leads to higher government income. Finally \textbf{(d)} shows that from the three taxation schemes proportional taxation achieves the highest average growth factor under an income maximizing government which can freely adjust the rate of administrative cost. The regressive scheme turns out to be too inefficient in turning on the portfolio re-balancing effect to enhance growth, while progressive taxation turns out to give incentives to raise the admin rate to such an extend that the loss due to this outweighs its efficiency in using the portfolio re-balancing effect.

By focusing on the portfolio re-balancing effect we proposed a new approach to think about the link between inequality, redistributive taxation and growth. With our simulation, we have shown that taxation and redistribution can be a crucial ingredient to ensure the survival and development of a society relying on risky multiplicative stochastic growth of human capital. Our approach gives another explanation of why redistribution can be beneficial for growth. Together with the other approaches mentioned in the introduction, we show that the interplay between inequality, redistribution and growth depends on preferences (fairness and insurance considerations), socio-political unrest (social stability), incentives and disincentives (effort), the calculus of the median voter (voting-system) and on the \emph{portfolio re-balancing effect}.

When paying taxes was voluntary, payment could be seen as an act of cooperation, which seems irrational but ensures the sustainable growth of human capital. As in the related public goods game, this society would be vulnerable to free-riders. In the public goods game the free rider problem is often solved by social norms or governmental forces to pay taxes.

At difference to the classical public goods game, where the public good is multiplied by an efficiency factor larger than one, our model does not have such an amplification. On the contrary, from the collected public good a fraction is subtracted for administrative costs, which is equivalent to an efficiency factor less than one. Consequently, the emergence of cooperation, i.e.~the sharing of income in order to sustain a long term growth, is even more subtle in our redistribution model, because even a normative call ``Pay your part and it will be immediately increased by an efficiency factor!'' does not work as easily.

If we assume that different societies compete, evolution would promote those societies with higher overall growth factors of their total income. Thus, there should be an evolutionary adaptation towards the optimal tax systems without assuming other forces. Such an idea is closely related to group selection as a mechanism to promote the evolution of cooperative behavior
\cite{Nowak2006FiveRulesEvolution}. Perhaps, the portfolio re-balancing effect is also a reason for the evolutionary success of religions which propose something like the tithe. Tithing 10\% of income to charity in a religious society ensures better growth of society's human capital which might be an evolutionary advantage against other societies because of the portfolio re-balancing effect.
The re-balancing effect might also be of relevance in other areas, such as biodiversity \cite{Schindler.Hilborn.ea2010Populationdiversityand} or knowledge sharing, to enhance innovativeness in social and economic systems.

Can we draw conclusions for large societies of some millions as in real world societies? We speculate that sizes larger than $N=1000$ imply even lower optimal tax rates because the portfolio re-balancing effect works even with very low tax rates and thus admin costs can be saved by low tax rates. But on the other hand we speculate that riskiness of individual stochastic growth also rises in larger societies which consequently implies higher optimal tax rates (see Figure \ref{fig:8}). It is much less likely to gain the best fitting human capital in a large society because there are more competitors. On the other hand, having gained the right human capital might lead to large benefits because there are many customers which could benefit from it. In conclusion, we speculate that our results are probably still valid for societies of real world sizes.

\section*{Acknowledgments}
JL thanks Wolfgang Breymann for pointing to literature and helpful discussions.

\bibliography{refs}


\begin{table}[!ht]
\caption{
\bf{Simulation setup.}}
\begin{center}
\begin{tabular}{ccc}
independent variable & range & size of range \\ \hline\hline
tax rate $a$ & $0, \stackrel{+0.02}{\dots}, 1$ & 51 \\
admin rate $b$ & $0, \stackrel{+0.02}{\dots}, 0.8$ & 41 \\
taxation scheme & regressive, proportional, progressive & 3 \\
$\stackrel{\mathrm{riskiness}}{(\mean{\eta},\mean{\eta}_\mathrm{geo})}$ & $\stackrel{\mathrm{less\ risky}}{(1.25,0.8)}$,$\stackrel{\mathrm{intermediate}}{(1.5,0.667)}$,$\stackrel{\mathrm{more\ risky}}{(3,0.333)}$ & 3 \\
number of agents $N$ & 10, 100 & 2 \\
$t_{\max}$ & 500 & 1 \\ \hline
\end{tabular}\end{center}
\begin{flushleft}Numbers of parameter values multiply to 37,638 combinations. 100 simulation runs with $h_i(0)=1$ were computed for each. Consequently, the growth factor $g$ was estimated regressing $\log(g)$ in $\log Y(t) = \log N + t\cdot\log(g)$ (notice the intercept is naturally fixed at $\log N$). For these 3,763,800 values of $g$ the geometric mean was computed for each combination over all 100 runs giving 37,638 \emph{average growth factors} as the basis for Figures \ref{fig:3}--\ref{fig:8}.
\end{flushleft}
\label{tab:simsetup}
\end{table}

\pagebreak
\mbox{}
\pagebreak

\section*{Supporting Material}

\texttt{matlab}-code to run simulation and produce figures:
\tiny
\begin{verbatim}
 function figs

% Produces all simulation data and all figures for the Paper 
% by Jan Lorenz, Fabian Paetzel and Frank Schweitzer, 2012
% programmed by Jan Lorenz 

% Fig 1; 4 subfigures: figExRedis figExRedisdynfee figExRedisproptax figExRedisdynmax
exampleRedistribution 

% Fig 2; 4 subfigures: figMSPgrowthExMain figMSPgrowthExSub1 figMSPgrowthExSub2 figMSPgrowthExSub3
exampleTrajectories

% Generate systematic simulation data
dataMSPgrowthrates % WARNING TAKES TIME! PROBABLY AT LEAST A WEEK ...

% Figs 3--8; 
colmap % save the colormap skewjet to file 
figColormap % figure of colormap
figSimulation % calls subfunctions plotfigs_fixedN, plotfigs_taxfunction and produces lots of subfigures


% FUNCTIONS PRODUCING FIGURES

function exampleRedistribution

figure(1);clf
y = [100 300 600 1000 1500 2100]';
a = 1/3;
b = 0.25;
f = getfee(y,a);
m = getmax(y,a);
ylim([0 1.1*max(y)]);xlim([0 length(y)+1]);hold on
xfill=[0.7 0.7 1.3 1.7 2.3 2.7 3.3 3.7 4.3 4.7 5.3 5.7 6.3 6.3];
yfill=a*[0 y(1) y(1) y(2) y(2) y(3) y(3) y(4) y(4) y(5) y(5) y(6) y(6) 0];
fill(xfill,yfill,[0.5 0.5 1],'EdgeColor',[0 0 1],'LineWidth',2,'FaceAlpha',0.5);
fill([0 length(y)+1 length(y)+1 0],[m m max(ylim) max(ylim)],[1 0.5 0.5],'LineStyle','none','FaceAlpha',0.5);
fill([0 length(y)+1 length(y)+1 0],[f f 0 0],[0.5 1 0.5],'LineStyle','none','FaceAlpha',0.5);
plot([0 length(y)+1],[m m],'r','Linewidth',2);
plot([0 length(y)+1],[f f],'g','Linewidth',2);
bar(y,'FaceColor',0.25*[1 1 1],'BarWidth',0.6);hold on;
box on
set(gca,'Layer','top','Xtick',1:length(y),'Position',[ 0.1600    0.1500    0.7750    0.8150])
make_figures('figExRedis',[82   192   215   332],'removeEps',1)
mode = {'proptax','dynfee','dynmax'};
for k=1:3
    afterRedis(k+1,mode{k},y,a,b)
    make_figures(['figExRedis' mode{k}],[5   243*(k-1)   0.6*[269   221]],'removeEps',1)
end


function exampleTrajectories
tmax=500;
% parameters for 4 figures
n = [10 10 10 100]; 
tax = [0.3 0.3 0.01 0.3];
adm = [0.2 0.6 0.2  0.2];
% log-normal
Mean = 1.5; geoMean = 1/Mean;
mu = log(geoMean); si = sqrt(2*(log(Mean)-mu));
r = exp(mu + si*randn(n(end),tmax));
maxrate=Mean*(1-adm.*tax);
for i=1:length(n)
    figure(i);clf
    x = ones(n(i),tmax);xcol='g';
    y = ones(n(i),tmax);ycol='b';
    z = ones(n(i),tmax);zcol='r';
    for t = 2:tmax
        x(:,t) = redis(prod(HCprod(x(:,t-1),'randomvector',r(1:n(i),t))),'dynfee','tax',tax(i),'adm',adm(i));
        y(:,t) = redis(prod(HCprod(y(:,t-1),'randomvector',r(1:n(i),t))),'proptax','tax',tax(i),'adm',adm(i));
        z(:,t) = redis(prod(HCprod(z(:,t-1),'randomvector',r(1:n(i),t))),'dynmax','tax',tax(i),'adm',adm(i));
    end
    plot(sum(x)',xcol);set(gca,'YScale','log');hold on
    plot(sum(y)',ycol);set(gca,'YScale','log')
    plot(sum(z)',zcol);set(gca,'YScale','log')
    plot(n(i)*maxrate(i).^(1:tmax),'k');
    pl = n(i)*geoMean.^(1:tmax);plot(pl(pl>0),'k');
    if i==1
        set(gca,'Position',[0.14 0.1 0.65 0.8])
        text(1.01*tmax,sum(x(:,end)),'regr. tax','color',xcol)
        text(1.01*tmax,sum(y(:,end)),'prop. tax','color',ycol)
        text(1.01*tmax,sum(z(:,end)),'prog. tax','color',zcol)
        text(1.01*tmax,(n(i)*maxrate(i)^tmax),'N((1-ba)<\eta>)^t','color','black')
        xlabel('t')
        ylabel('total income (\Sigma_i y_i(t))');
    end
    switch i
        case 1
            title(['N = ' num2str(n(i)) ', a = ' num2str(tax(i)) ', b = ' num2str(adm(i))]);
        case 2
            title(['N = ' num2str(n(i)) ', a = ' num2str(tax(i)) ', {\bf b = ' num2str(adm(i)) '}']);
        case 3
            title(['N = ' num2str(n(i)) ', {\bf a = ' num2str(tax(i)) '}, b = ' num2str(adm(i))]);
       case 4
            title(['{\bf N = ' num2str(n(i)) '}, a = ' num2str(tax(i)) ', b = ' num2str(adm(i))]);
    end
    xlim([0 500])
    ylim([min([sum(x) sum(y) sum(z)])*10^-10 max(n(i)*maxrate(i)^tmax)^1.05])
    if i==1
        g=axes('position',[0.19 0.65 0.17 0.15]);
        X = 0:0.01:5;plot(X,lognpdf(X,mu,si),'k');
        set(g,'XTick',[0 1 2],'YTick',[],'YColor','w','Clipping','off');
        ylim([0 0.9*max(ylim)])
        box off
        text(0.8,0.9*max(ylim),['\eta log-normal']) 
        text(1.2,0.6*max(ylim),['\mu=' num2str(round(mu*1000)/1000) ', \sigma=' num2str(round(si*1000)/1000)])
        text(1.7,0.3*max(ylim),['<\eta> = ' num2str(round(Mean*1000)/1000) ', <\eta>_{geo} = ' num2str(round(geoMean*1000)/1000)])
        title('HCprod, pdf \eta')
    end
    switch i
        case 1
            make_figures('figMSPgrowthExMain',[877   373   400   350]);
        case 2
            make_figures('figMSPgrowthExSub1',[600   173   [400   350]/2.3]);
        case 3
            make_figures('figMSPgrowthExSub2',[800   173   [400   350]/2.3]);
        case 4
            make_figures('figMSPgrowthExSub3',[1000   173   [400   350]/2.3]);
    end
end

function afterRedis(k,mode,y,a,b)
figure(k);clf;
pg=sum(y)*a*(1-b);
switch mode
    case 'proptax';lcol = [0.5 0.5 1];fcol = [0 0 1];
    case 'dynfee';lcol = [0.5 1 0.5];fcol = [0 1 0];
    case 'dynmax';lcol = [1 0.5 0.5];fcol = [1 0 0];
end
bar(1:length(y),redis(y,mode,'adm',b,'tax',a),'FaceColor',fcol,'BarWidth',0.5); hold on
bar(1:length(y),ones(1,length(y))*pg/length(y),'FaceColor',lcol,'BarWidth',0.5);
set(gca,'Layer','top','Xtick',1:length(y),'Position',[ 0.200    0.1500    0.7750    0.8150])
box on;ylim([0 1.1*max(y)]);xlim([0 length(y)+1]);


function figColormap
figure(1);clf
load colmap
c = exp([-1.5 1]);caxis(c);
imagesc(linspace(c(1),c(2)),1,linspace(c(1),c(2)));
colormap(skewjet)
set(gca,'YTick',[],'XAxisLocation','top','Position',[0.1 0.35 0.8 0.15])
text(0.99,2.5,'| growth \rightarrow')
text(0.99,2.5,'\leftarrow destruction ','HorizontalAlignment','right')
title({'color axis: average growth factor g(b,a)'})
make_figures('figColormap',[508   500   313   100])

function figSimulation
mean_eta = [1.25, 1.5, 3]; geomean_eta = 1./mean_eta;
MU = log(geomean_eta); SI = sqrt(2*(log(mean_eta)-MU));
for n = [10 100]
    for dist_ind = [1 2 3]
        mu = round(100*MU(dist_ind)); si = round(100*SI(dist_ind));
        plotfigs_fixedN(n,mu,si)
    end
end
modes = {'dynfee','proptax','dynmax'};
for mode = 1:3
    for dist_ind = [1 2 3]
        mu = round(100*MU(dist_ind)); si = round(100*SI(dist_ind));
        NN = [10 100];
        plotfigs_taxfunction(modes{mode},mu,si,NN)
    end
end

function plotfigs_taxfunction(taxfunction,muName,siName,NN)
modes = []; mu=[]; % necessary initialization
figure(1);clf;
for nInd = 1:length(NN)
    load(['data' 'N' num2str(NN(nInd)) 'mu' num2str(muName) 'si' num2str(siName) '.mat'])
    mode = find(strcmp(taxfunction,modes));
    if length(NN) == 2
        linecol = ones(2,3)-[0.33*[1 1 1]; [1 1 1] ];
    else
        linecol = ones(3,3)-[0.33*[1 1 1]; 0.66*[1 1 1]; [1 1 1] ];
    end
    switch mode
        case 1; linecol(:,2)=1;
        case 2; linecol(:,3)=1;
        case 3; linecol(:,1)=1;
    end
    g=growrate(1,1,1,:);g=g(:);growrate=growrate(:,:,:,~isnan(g));find(~isnan(g),1,'last')
    P = mean(growrate(:,:,mode,:),4);
    [tmp,maxind]=max(P);argmaxA = A(maxind);
    [C(nInd).h,h(nInd).h]=contour(B,A,P,'color',linecol(nInd,:),'linewidth',2,'LevelList',0); hold on
    T(nInd).T=clabel(C(nInd).h,h(nInd).h,'LabelSpacing',500,'BackgroundColor',[1 1 1]);
    set(T(nInd).T,'string',num2str(NN(nInd)));
    plot(B,smooth(argmaxA),'--','color',[linecol(nInd,:)]);
end
figure(1);
axis equal;axis([min(B) max(B) min(A) max(A)])
switch mode
    case 1; titlestr = 'regressive tax';
    case 2; titlestr = 'proportional tax';
    case 3; titlestr = 'progressive tax';
end
plot(B,1./B*(1-1/exp(mu+si^2/2)),'k:');
make_figures(['figSim' taxfunction 'mu' num2str(muName) 'si' num2str(siName)],[360   530   233   193])

function plotfigs_fixedN(n,mu,si)
load(['data' 'N' num2str(n) 'mu' num2str(mu) 'si' num2str(si) '.mat'])
g=growrate(1,1,1,:);g=g(:);growrate=growrate(:,:,:,~isnan(g));find(~isnan(g),1,'last')
figure(1);clf;figure(2);clf;figure(3);clf;
linespec={'g','b','r'};
maxadminIncomeInd = zeros(1,3);
maxmaxP = zeros(1,3);
for mode=1:3
    P = exp(mean(growrate(:,:,mode,:),4)); % mean(growrate(:,:,mode,:),4) is log of growthrate of model y(t)=N*exp(...)^t
    % P is thus geometric mean of growthrates y(t) = N*P^t
    [maxP,maxind]=max(P);argmaxA = A(maxind);
    smoothargmaxA=smooth(argmaxA)';
    adminIncome = smooth(maxP.*smoothargmaxA.*B)';
    [tmp,maxadminIncomeInd(mode)] = max(adminIncome);
    maxmaxP(mode) = maxP(maxadminIncomeInd(mode));
    figure(1)
    subplot(1,3,mode)
    load('colmap'); % load the colormap skewjet, can be replaced by a default one
    colormap(skewjet)
    imagesc(B,A,P);axis xy;axis equal;axis tight;
    hold on;
    contour(B,A,P,[1 1],'color','k')
    plot(B,smoothargmaxA,'k--')
    c = exp([-1.5 1]);caxis(c);
    if mode==1;ylabel('tax rate a');end; if mode==2;xlabel('admin rate b');end
    switch mode
        case 1; title('{\bf regressive tax}','color','g');
        case 2; title('{\bf proportional tax}','color','b');
        case 3; title('{\bf progressive tax}','color','r');
    end
    plot(B,1./B*(1-1/exp(mu+si^2/2)),'k:');
    figure(2);
    contour(B,A,P,[1 1],'color',linespec{mode},'linewidth',2); hold on
    C(mode)=plot([100],[100],linespec{mode},'linewidth',2); % just for the legend
    plot(B,smoothargmaxA,[linespec{mode} '--']);
    figure(3);    
    subplot(3,1,1)
    plot(B,smoothargmaxA,[linespec{mode} '--']);hold on
    plot(B(maxadminIncomeInd(mode)),smoothargmaxA(maxadminIncomeInd(mode)),[linespec{mode} '*'])
    subplot(3,1,2)
    plot([0 0.8],[1 1],'color',0.5*[1 1 1]);hold on
    plot(B,maxP,'color',linespec{mode});
    plot(B(maxadminIncomeInd(mode)),maxP(maxadminIncomeInd(mode)),[linespec{mode} '*'])
    subplot(3,1,3)        
    plot(B,adminIncome,'color',linespec{mode});hold on
    plot(B(maxadminIncomeInd(mode)),adminIncome(maxadminIncomeInd(mode)),[linespec{mode} '*'])
end
figure(1)
make_figures(['fig' 'Sim' '1' 'N' num2str(n) 'mu' num2str(round(mu*100)) 'si' num2str(round(si*100))],[6   533   522   188])
figure(2);
axis equal;axis([min(B) max(B) min(A) max(A)])
plot(B,1./B*(1-1/exp(mu+si^2/2)),'k:');
ylabel('tax rate a');xlabel('admin rate b');
make_figures(['fig' 'Sim' '2' 'N' num2str(n) 'mu' num2str(round(mu*100)) 'si' num2str(round(si*100))],[985   435   292   288])
figure(3)
subplot(3,1,1); %title({'optimal tax rate'})
ylabel('a_{opt}');xlim([0 0.8]);ylim([0 1])
subplot(3,1,2); %title('maximal growth factor')
ylabel('g_{max}');axis tight;yax=ylim;xlim([0 0.8]);ylim([0.9*yax(1) 1.05*yax(2)])
[tmp,i] = max(maxmaxP);
plot([0 0.8],max(maxmaxP)*[1 1],[linespec{i} '--'])
subplot(3,1,3); %title('government income')    
xlabel('admin rate b');ylabel('gov_{opttax}');axis tight;yax=ylim;xlim([0 0.8]);ylim([0 1.05*yax(2)])
make_figures(['fig' 'Sim' '3' 'N' num2str(n) 'mu' num2str(round(mu*100)) 'si' num2str(round(si*100))],[808   156   213   634])


% FUNCTIONS PRODUCING DATA

function dataMSPgrowthrates
% makes samplesize=100 simulation runs for each data point for different
% values of mu, sigma, N, and taxregime; and estimate a growthrate for
% each; writes this data to files to be read be figure-producing functions
mean_eta = [1.25, 1.5, 3]; geomean_eta = 1./mean_eta;
MU = log(geomean_eta); SI = sqrt(2*(log(mean_eta)-MU));
N = [10 100];
for n = N
    for dist_ind = 1:length(mean_eta)
        mu = MU(dist_ind); si = SI(dist_ind);
        files = dir;
        if ~any(strcmp({files.name},['data' 'N' num2str(n) 'mu' num2str(round(mu*100)) 'si' num2str(round(si*100)) '.mat']))
            % external parameter
            tmax=500;
            samplesize = 100;
            % running parameters
            A = 0:0.02:1;
            B = 0:0.02:0.8;
            modes = {'dynfee','proptax','dynmax'};
            y = ones(n,tmax);
            growrate = nan(length(A),length(B),length(modes),samplesize);
            SSE = nan(length(A),length(B),length(modes),samplesize);
            RMSE = nan(length(A),length(B),length(modes),samplesize);
            RSQUARE = nan(length(A),length(B),length(modes),samplesize);
        else % when a file exists one might resume and finish an old computation
            tmax = 500;
            samplesize = 100;
            y = ones(n,tmax);
            load(['data' 'N' num2str(n) 'mu' num2str(round(mu*100)) 'si' num2str(round(si*100)) '.mat'])
        end
        for s=1:samplesize
            r = exp(mu + si*randn(n,tmax));
            for a=1:length(A)
                for b=1:length(B)
                    for m=1:length(modes)
                        if isnan(growrate(a,b,m,s))
                            y(:,1) = 1;
                            for t = 2:tmax
                                 y(:,t) = redis(prod(HCprod(y(:,t-1),'randomvector',r(:,t))),modes{m},'tax',A(a),'adm',B(b));
                            end
                            % fit for poly1: ly = p1*ly + p2
                            ly = log(sum(y))';
                            if all(~isinf(abs(ly)))
                                [f,gof] = fit((0:tmax-1)',ly,'poly1','Lower',[-inf log(n)],'Upper',[inf log(n)]);
                                growrate(a,b,m,s) = f.p1;
                                SSE(a,b,m,s) = gof.sse;
                                RMSE(a,b,m,s) = gof.rmse;
                                RSQUARE(a,b,m,s) = gof.rsquare;
                            end
                        end
                    end
                end
            end
            save(['data' 'N' num2str(n) 'mu' num2str(round(mu*100)) 'si' num2str(round(si*100)) '.mat'],'A','B','RMSE','RSQUARE','SSE','growrate','modes','mu','si','tmax')
        end
    end
end


% FUNCTIONS FOR COMPUTATION


function y = prod(h,varargin)
y = h;

function h = HCprod(y,varargin)
% y is a row vector of the current wealth of n=length(y) agents 
% defaults and varargins
r = [];
for i=1:length(varargin)/2
    switch varargin{i*2-1}
        case 'randomvector';r = varargin{i*2};
    end
end
% production mechanism
h = y.*r;

function yn = redis(y,mode,varargin)
% defaults and varargins
tax = [];
adm = [];
for i=1:length(varargin)/2
    switch varargin{i*2-1}
        case 'tax';tax = varargin{i*2};
        case 'adm';adm = varargin{i*2};
    end
end
% redistribution mechanism
switch mode
    case 'proptax'
        % public good, prop tax, admin cost
        if isempty(tax); tax = 0.5; end % tax rate for the public good 
        if isempty(adm); adm = 0; end % rate of administrative cost for public good
        t = sum(y*tax);
        pg = sum(t)*(1-adm);
        yn = y*(1-tax)+pg/length(y);
    case 'dynfee'
        % public good, a fixed amount is taxed (or all)
        if isempty(tax); tax = 0.5; end % tax rate, what proportion to get from total wealth 
        if isempty(adm); adm = 0; end % rate of administrative cost for public good
        f = getfee(y,tax);
        t = min([y f*ones(size(y))],[],2);
        pg = sum(t)*(1-adm);
        yn = (y-t)+pg/length(y);
    case 'dynmax'
        % public good, all over max is gathered for pg, admin cost
        if isempty(tax); tax = 0.5; end % tax rate, what proportion to get from total wealth 
        if isempty(adm); adm = 0; end % rate of administrative cost for public good
        m = getmax(y,tax);
        t = max([zeros(size(y)) y-m],[],2);
        pg = sum(t)*(1-adm);
        yn = (y-t)+pg/length(y);
end

function f = getfee(y,tax)
% y is row vector of numbers, tax is a desired tax rate
% extracts f such that sum(max([zeros(size(y) y-m],[],2))= sum(y)*tax
y = [0 ; sort(y,'ascend')];
t = sum(y)*tax;
n = length(y);
i = 1;
while sum(y(1:i-1))+y(i)*(n-i+1)<t 
    i=i+1;
end
d = sum(y(1:i-1))+y(i)*(n-i+1)-t;
f = y(i) - d/(n-i+1);

function m = getmax(y,tax)
% y is row vector of numbers, tax is a desired tax rate
% extracts m such that sum(max([zeros(size(y) y-m],[],2))= sum(y)*tax
y = [sort(y,'descend') ; 0];
t = sum(y)*tax;
i = 1;
while sum(y(1:i)-y(i+1))<t % optimizing potential for large y!
    i=i+1;
end
d = sum(y(1:i)-y(i+1))-t;
m = y(i+1) + d/i;

% MISCELLANEOUS FUNCTIONS FOR PRINTING AND COLORMAPS 

function make_figures(savestr,pos,varargin)
% defaults removeEps = 0, pdfcrop = 0
removeEps = 0;
pdfcrop = 0;
if mod(length(varargin),2)==1
    error('varargin not correct');
else
    for i=1:length(varargin)/2
        switch varargin{i}
            case 'removeEps'
                removeEps = varargin{i+1};
            case 'pdfcrop'
                pdfcrop = varargin{i+1};
        end
    end
end
set(gcf,'PaperPositionMode','auto','Position',pos)
print(savestr,'-deps2c')
eval(['! epstopdf ' savestr '.eps'])
if removeEps==1
    eval(['!rm ' savestr '.eps'])
end
if pdfcrop==1
    eval(['!pdfcrop ' savestr '.pdf ' savestr '.pdf'])
end


function colmap
% Code ommitted. Just saves a particular color map as a file to load it for graphics
\end{verbatim}

\end{document}